\documentstyle[11pt,newpasp,twoside]{article}
\def\edcomment#1{\iffalse\marginpar{\raggedright\sl#1\/}\else\relax\fi}
\reversemarginpar

\begin{document}
\title{The Fundamental Properties of Dwarf Elliptical Galaxies in Clusters}
 \author{Rafael Guzm\'an, Alister W. Graham, Ana Matkovi\'c
\& Ileana Vass}
\affil{Department of Astronomy, Bryant Space Science Center, University of Florida, Gainesville, FL 32611}
\author{Javier Gorgas \& Nicolas Cardiel}
\affil{Facultad de Ciencias Fisicas, Universidad Complutense de Madrid, 28040 
Madrid, Spain}

\begin{abstract}
We present preliminary results of an extensive study of the
fundamental properties of dwarf elliptical galaxies (dEs) in the Coma
cluster. Our study will combine HST surface photometry with
ground-based UBRIJK photometry and optical spectroscopy. The combined
data set will be used to investigate the intrinsic correlations among
global parameters in cluster dEs, including the Fundamental Plane,
color-magnitude, Faber-Jackson's, Kormendy's, and velocity dispersion
vs. line strength indices. These empirical correlations have provided
important constraints to theoretical models of galaxy formation and
evolution for ``normal'' ellipticals. Although dEs are the most
abundant galaxy population in clusters their properties remain,
however, largely unknown. Our study aims to provide an essential
reference for testing current theories on the formation and evolution
of dEs in clusters, and understanding their relation to more massive
ellipticals.
\end{abstract}

\section{Dwarf Ellipticals: a Key Player in Galaxy Evolution?}

	Dwarf elliptical galaxies (dEs) are low-mass, spheroidal-like,
quiescent galaxies with $M_B > -18$ (see Ferguson \& Binggeli 1994 for a
comprehensive review). In the last decade, there has been an increased
interest in dEs driven by two main results.

	Firstly, there seems to be an apparent dichotomy in some
empirical correlations among global galaxy parameters of dEs and
normal ellipticals. For elliptical galaxies, these empirical
scaling-laws between size, luminosity, surface brightness, color,
velocity dispersion, and line strength indices provided: (i) a
characterization of the structural, kinematic, and stellar
population properties of this galaxy class; (ii) constraints to
theoretical models of galaxy formation and evolution of ellipticals;
and (iii) distance indicators to map the local peculiar velocity field
(Terlevich et al.\ 1981, Djorgovski \& Davis 1987; Guzm\'an et
al.\  1993). Initial studies of Local Group galaxies showed that dEs
exhibit different correlations (e.g., luminosity vs. surface
brightness) suggesting different physical processes during formation
and/or evolution (e.g., Wirth \& Gallagher 1984; Kormendy 1985;
Bender, Burstein \& Faber 1992).

	Secondly, recent theoretical models of galaxy formation and
evolution have highlighted the potential importance of dEs in modern
cosmology. For instance, in field galaxy studies, dEs have been
proposed as the faded counterparts of the ubiquitous low-mass
starbursting galaxies at redshifts $z<1$ whose number density greatly
exceeds the non-evolution model predictions (the so-called ``faint
blue galaxies'', Babul \& Rees 1992; Babul \& Ferguson 1996). In
cluster galaxies studies, dEs have been proposed to be the evolved
counterparts of the numerous low-mass blue disk galaxies observed in
clusters at $z\sim$0.5 (e.g., the so-called ``galaxy harassment''
scenario; Moore, Lake \& Katz 1998). In particular, the galaxy
harassment model makes very specific predictions on the number
density, kinematics, and stellar population properties of dEs located
in the inner and outer areas of clusters which can be directly tested
with observations.

	Despite their potential cosmological importance, and the fact
that they are the most numerous galaxy type in the local universe ---at
least in clusters---, dEs are among the most poorly studied due to their
characteristic low surface brightness (22 $<$ SBe $<$ 26 B-mag/arcsec$^2$).
A summary of the current sample size of dEs for which various photometric 
and spectroscopic properties have been studied is given below:

\begin{itemize}
\item integrated photometry (e.g., colors and magnitudes): $\sim10^3$ dEs
mainly in Coma, Virgo, and Fornax clusters (e.g., Binggeli \& Cameron
1991; Vader \& Chaboyer 1994; Ulmer et al.\  1996, Secker, Harris \&
Plummer 1997; Terlevich, Caldwell \& Bower 2001).

\item surface photometry (e.g., size, surface brightness, and luminosity
profile): $\sim10^2$ dEs mainly in the Virgo and Coma clusters (e.g.,
Binggeli \& Cameron 1991; Ferguson 1992; Young \& Currie 1998).
 
\item line strength indices (e.g., H$\beta$, Mg$_2$, and Fe): $\sim10^2$ dEs in
Coma, Virgo, Fornax and the Local Group (e.g., Brodie \& Huchra 1991;
Held \& Mould 1994; Gorgas et al.\  1997; Secker et al.\  1998; Mobasher
et al.\  2002).

\item internal kinematics (e.g., velocity dispersion, and rotation): $\sim10$
dEs in Virgo and the Local Group (e.g., Bender \& Nieto 1990; Peterson
\& Caldwell 1993; Pedraz et al.\  2001; Simien \& Prugniel 2002; Geha et
al.\  2002).

\end{itemize}

We are currently conducting en extensive study of $\sim$100 dEs
located in the inner and outer regions of the Coma cluster using
HST/WFPC2(ACS) and WIYN/HYDRA. This study aims to provide the first
large, homogeneous data set of photometric and spectroscopic
parameters of dEs outside the Local Group, including luminosity, size,
surface brightness, {\sl velocity dispersion}, mass-to-light ratio,
optical and infrared colors, and line strength indices. Our goals are:
(i) to investigate the relation between dEs and elliptical galaxies in
clusters; and (ii) to test galaxy harassment model predictions on the
structural, stellar population, and kinematic properties of cluster
dEs, and their dependence on the environment.

\section{The Data}

\subsection{Sample Selection}

	Coma cluster dE-candidates were chosen using U, B, and R
images obtained with WIYN/MiniMo and INT/Wide Field Camera. The
criteria used for the sample selection can be summarized as: (i) $0.2
< (U-B) < 0.6$, and $1.3 < (B-R) < 1.5$ in order to minimize
contamination from background field disk galaxies at z$\sim$0.2; and
(ii) $17.5 < B < 20.5$ to select galaxies with $-18 < M_B < -15$,
approximately at the distance of the Coma cluster. Figure 1
illustrates these criteria. A detailed description is provided in
Matkovi\'c \& Guzm\'an (in prep).

\begin{figure}
\vspace{5cm}
\includegraphics{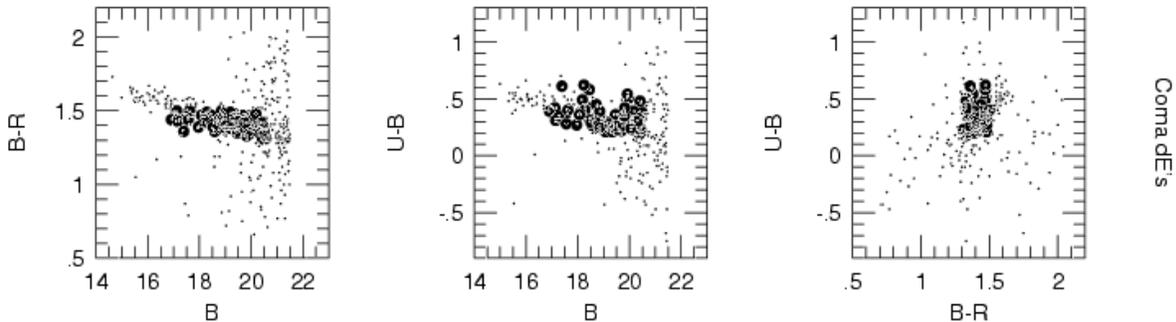}
\caption{Selection of dE candidates in the Coma cluster (solid circles) following the criterion described in the text.}
\end{figure}

	The projected surface density of dE-candidates in the inner
region (1$^{\circ}$ from the cluster center) is $\sim$700 per square
degree.  In the outer region (between 1$^{\circ}$ and 2.5$^{\circ}$
from the cluster centre) the projected surface density is $\sim$100
per square degree. These densities are ideal for large FOV multiobject
spectrographs, such as WIYN/HYDRA. The selection criteria turned out
to be remarkably reliable: about 90\% of dE-candidates observed with
HYDRA are spectroscopically confirmed to be cluster members.

\subsection{WIYN/HYDRA Spectroscopy}

Spectra of $\sim$50 dEs and $\sim$20 Es in the Coma cluster region
were obtained using HYDRA at the WIYN telescope. The instrumental
setup adopted for these observations provided a spectral resolution of
FWHM = 130 km/s with a 3 arcsec fiber diameter. The spectral
rest-frame wavelength ensured observations of strong absorption
features such as H$\delta$, G-band, H$\gamma$, H$\beta$, Mg$_2$, and
Fe$\lambda$5350.  The total integration time ranged from 1hr to 9hr,
depending on the galaxy magnitude. Figure 2a shows a representative
example of WYIN/HYDRA spectra.  Velocity dispersion ($\sigma$)
measurements were measured using the Fourier quotient method as
described by Gonz\'alez (1994). The values measured for our dE and E
samples range from 30 to 100 km/s. A full description of the
spectroscopic measurements and the data reduction is provided in
Matkovi\'c \& Guzm\'an (in prep).


\subsection{HST Imaging}

About 60 archival HST/WFPC2 F606W images of 20 different fields in the
inner region of the Coma cluster were investigated. Eighteen Coma dE
candidates in our sample were identified in these HST images. The
images were reduced following the standard HST pipeline. The most
difficult step in the data reduction is to ensure an accurate
background subtraction since many dE galaxies are affected by the
large, extended halos of more massive cluster galaxies. This was
achieved using a wavelet decomposition method that optimally removes
any large-scale structure in the background due to the contamination
of nearby galaxy halos. Figure 2b shows a panel with 4 HST F606W
images of dE galaxies in Coma. For a full account of the data
reduction procedure see Graham \& Guzm\'an (2002). 

\begin{figure}
\vspace{7cm}
\includegraphics{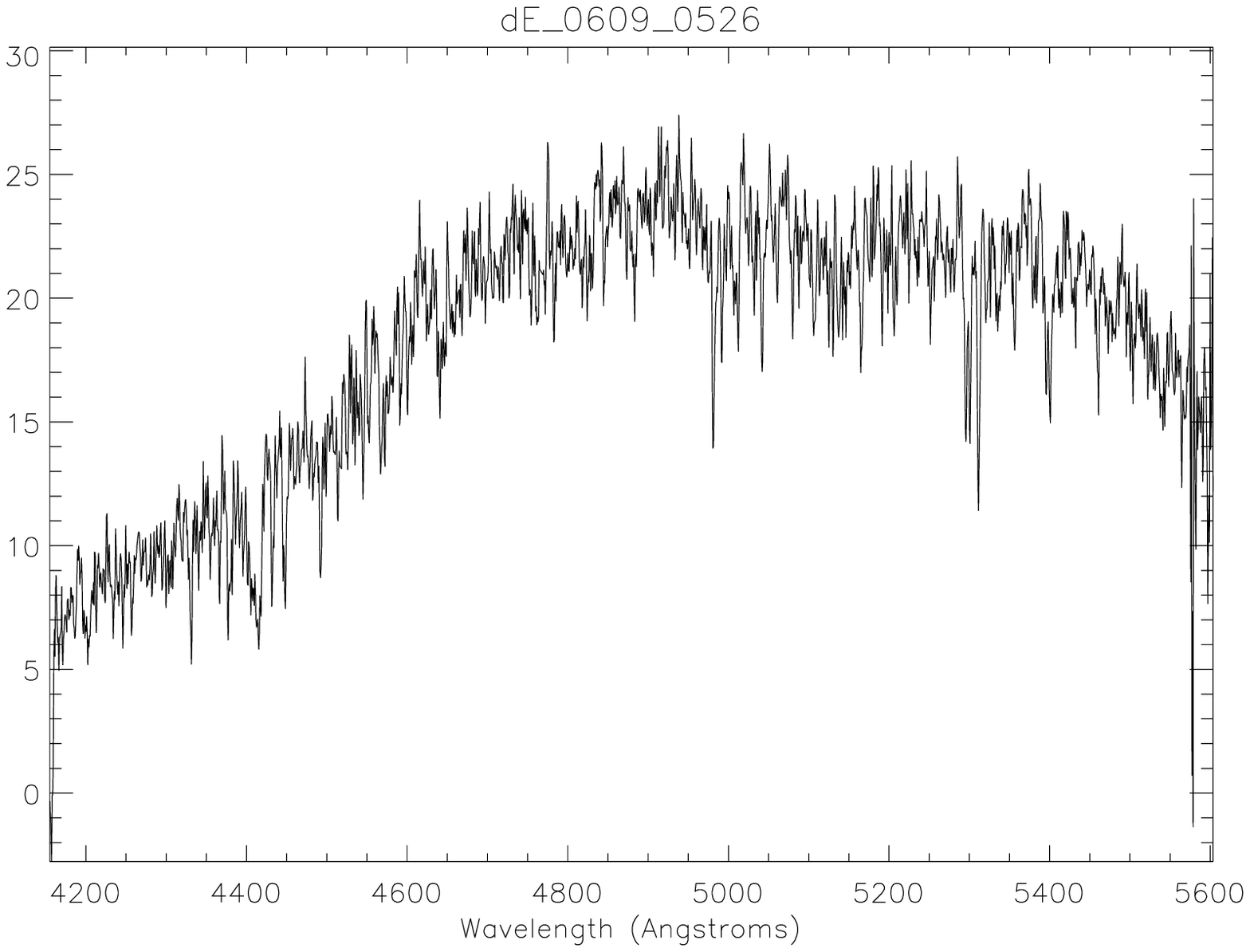}
\includegraphics{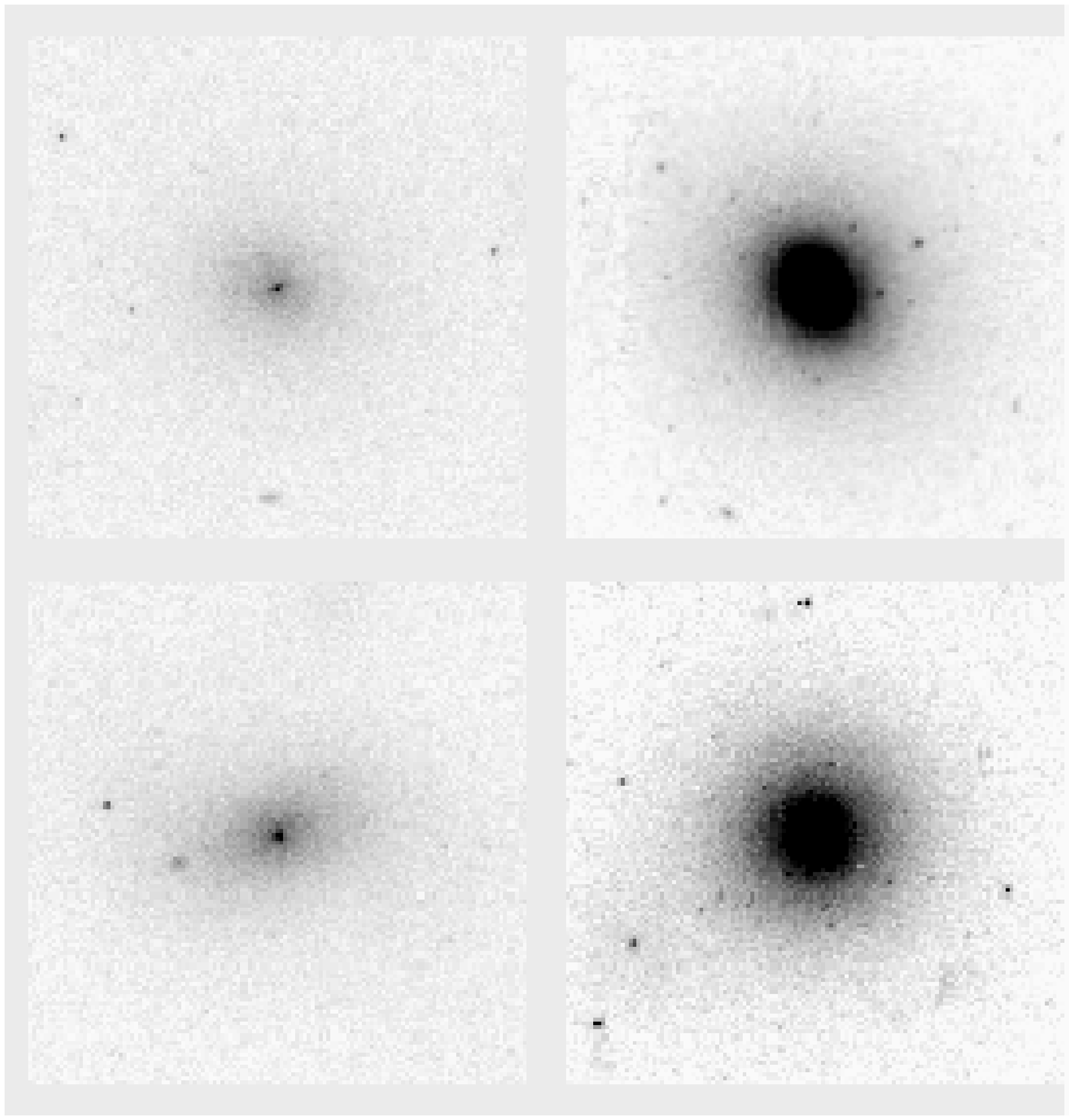}
\caption{(a) WIYN/HYDRA spectrum of a Coma dE galaxy. (b) HST/WCPC2 images
of four dE galaxies in Coma.}
\end{figure}

	Photometry was performed using the IRAF task ELLIPSE and the
resulting surface brightness profiles were fitted with a combination of
a PSF-convolved S\'ersic model and either a point source or a
PSF-convolved central Gaussian (Graham \& Guzm\'an 2002). This
provided measurements of the half-light radius ($R_e$), the effective
surface brightness ($\mu_e$), the S\'ersic index ($n$), and the
total galaxy magnitude. Nucleation was detected in all but two of our
final 15 dE galaxy sample. Our error estimates include the effects of:
shot-noise, placement on the chip, estimation of the galaxy
orientation and center, and the associated influence of pixelation. In
all cases when we had multiple light-profiles, which we did for most
galaxies, the surface brightness profiles were consistent with each
other down to a level of $\mu_{\rm F606W}\sim 25$ mag arcsec$^{-2}$.
Fitting models down to this surface brightness level we found the
differences in the best-fitting S\'ersic parameters obtained from
different profiles of the same galaxy spanned $\pm$0.05 mag
arcsec$^{-2}$ in $\mu_e$, $\pm$5\% in $R_e$, and $\pm$4\% in $n$.

Figure 3 shows the surface brightness profiles for 9 Coma dEs in our
sample.

\begin{figure}
\vspace{16cm}
\includegraphics{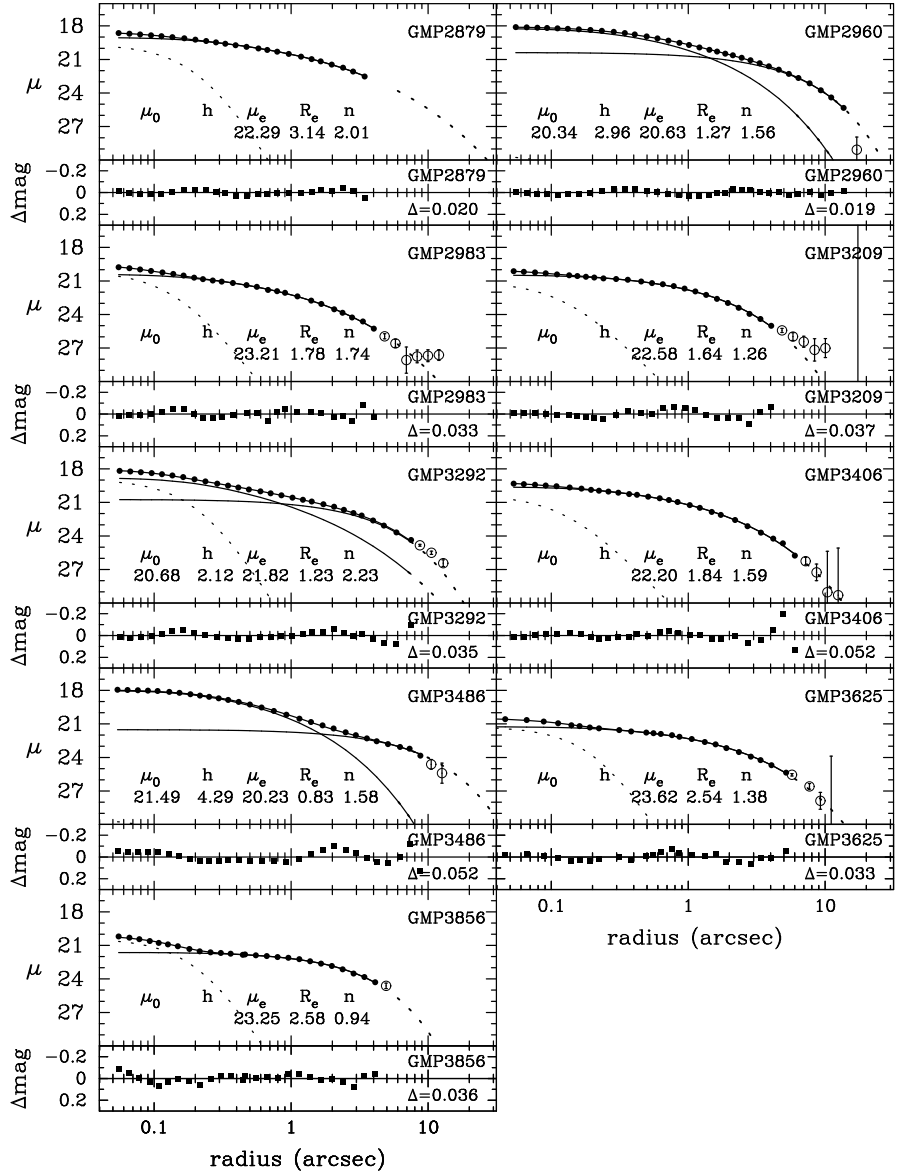}
\caption{Geometric mean-axis ($R=\sqrt{ab}$) 
surface brightness profiles for some of our Coma dEs.
Every profile has been fitted with a Moffat-convolved
S\'ersic model (solid line).  Three profiles are additionally fitted 
with an outer exponential (also a solid line).  
An inner point-source, when detected, is shown by a dotted line.  
The outer (extrapolated) model is also shown by a dotted line.  
Only the filled circles were used in the modelling process, the 
larger open circles were not.  
The lower panel displays the residuals of the data about the fitted 
model.  The mean residual from the fit is given by $\Delta$ mag.}
\end{figure}

\begin{figure}
\vspace{8.5cm}
\includegraphics{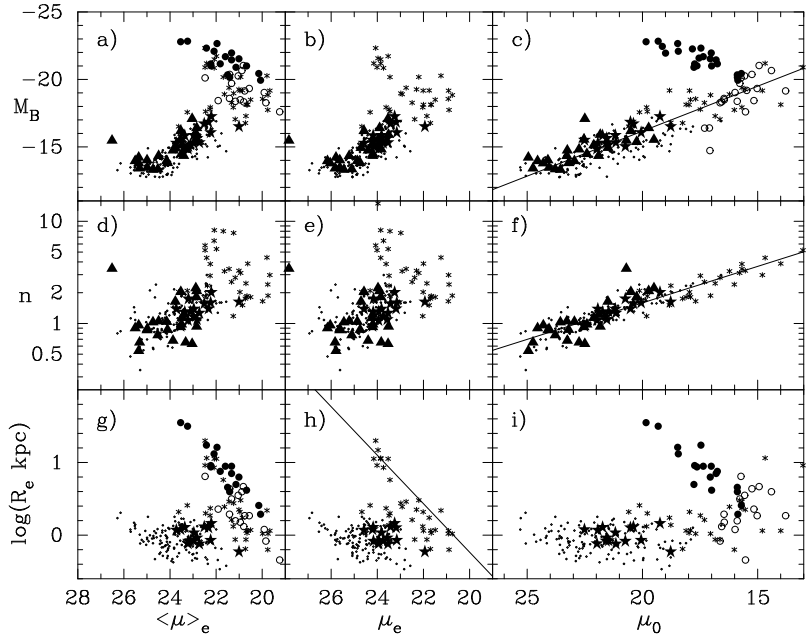}
\caption{The mean surface brightness within $R_e$ ($<\mu>_e$), the
surface brightness at $R_e$ ($\mu_e$), and the central host galaxy
surface brightness ($\mu_0$) are shown against the {\sl host} galaxy
magnitude ($M_B$), the global profile shape ($n$), and the half-light
radius ($R_e$).}
\end{figure}

\section{Is there a ``Dichotomy'' in the Structural Properties of dE 
and E Galaxies?}

The realization that dwarf ellipticals could be reasonably well
described with an exponential function (i.e., $n=1$) and that bright
ellipticals are better fit with de Vaucouleurs' $r^{1/4}$-law (i.e.,
$n=4$) led to the notion that they are two distinct families of
galaxies (Faber \& Lin 1983; Binggeli, Sandage \& Tarenghi 1984; Wirth
\& Gallagher 1984, but see Graham 2002).  One of the most referenced
papers to support this view is Kormendy (1985).  Plotting central
surface brightness ($\mu_o$) against luminosity, Kormendy showed two
relations almost at right angles to each other: one for the faint dE
galaxies and the other for the more luminous E galaxies. Similarly,
diagrams using the surface brightness at the effective half-light
radius ($\mu_e$) also show two somewhat perpendicular relations. These
diagrams are shown in Figure 4 for a large sample of 247 dE and E
galaxies. The symbols represent the following samples: dots correspond
to Binggeli \& Jerjen (1998) dE sample; triangles correspond to
Stiavelli et al.\ \ (2001) dE sample; large stars represent our Coma dE
galaxies; asterisks represent intermediate to bright E galaxies from
Caon et al.\ \ (1993); and open and solid circles represent the
so-called ``power-law'' and ``core'' E galaxies, respectively, from
Faber et al.\ \ (1997). Panels (a) (cf. Wirth \& Gallagher 1984) and (g)
(cf. Kormendy 1985) illustrate the two most widely accepted results
supporting the dichotomy between the structural properties of bright
Es and dEs.

\begin{figure}
\vspace{6cm}
\includegraphics{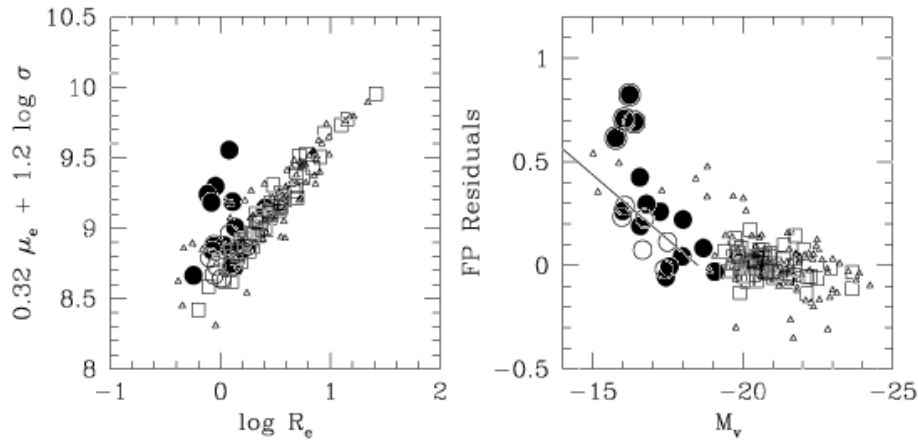}
\caption{(a) Edge-on view of the fundamental plane. Open squares: Coma
Es (Guzm\'an, 1994); open triangles: early-type galaxy sample (BBF,
1992); open circles: Virgo dEs (Geha et al 2002); solid circles: Coma
dEs (this work). (b) Residuals of the fundamental plane versus
absolute magnitude. Symbols as before. The solid line is not a fit to
the data but represents the expected trend for dEs in this diagram as
predicted by the galaxy properties framework described in Guzm\'an,
Lucey \& Bower (1993).}
\end{figure}

	The structural dichotomy is also apparent when studying the
fundamental plane (Djorgovski \& Davies 1987). Figure 5a shows the
edge-on projection of the fundamental plane for various samples of E
and dE galaxies. Our sample of 14 Coma cluster dEs used in this
preliminary study more than doubles the number of objects in previous
samples used for fundamental plane studies of dEs in the
literature. Clearly, dE galaxies lie off and above the relation
defined by ``normal'' Es (cf. Bender, Burstein \& Faber
1992). 

	In order to reasses critically the evidence for a physical
discontinuity in the structural properties of E and dEs galaxies we
note the following three important facts. Firstly, the observed
difference in the fundamental plane, $M_B-\langle\mu\rangle_e$, and
$R_e-\langle\mu\rangle_e$ diagrams {\sl are not independent results}
but simply {\sl three different ways to plot the same effect}. For
instance, the solid line plotted in Figure 5b is not a fit to the data
but the same trend defined by dEs in Figure 4a transformed using the
empirical framework of galaxy properties described by Guzm\'an, Lucey
\& Bower (1993).

	Secondly, in contrast to the above, there is substantial
evidence for a continuity, rather than a dichotomy, between the
alleged E and dE classes of galaxies. Caldwell (1983; their Figure 6)
showed that, fainter than $M_B\sim -20.5$, there is a continuous trend
between the central surface brightnesses and absolute magnitudes of dE
and E galaxies -- more luminous galaxies have brighter central surface
brightnesses (see also Caldwell 1987; Caldwell and Bothun 1987; Hilker
et al.\ \ 1999). Caldwell (1983) also revealed a continuous and linear
relationship exists between ($U-V$) color and luminosity over the
magnitude interval $-23 < M_V < -15$ (see also Terlevich, Caldwell, \&
Bower 2001). Caldwell \& Bothun (1987) further revealed a continuous
luminosity--metallicity relation across the alleged dE and E classes
(see also Terlevich et al.\ \ 1981). The relation between luminosity and
velocity dispersion for E galaxies (Faber \& Jackson 1976)
has also been shown to extend linearly to include the dE galaxies
(e.g., Bender, Burstein, \& Faber 1992). None of these correlations
with luminosity suggest evidence for a discontinuity at $M_B\sim -18$
which apparently denotes the transition between two distinct galaxy
classes (E and dE).

	Thirdly, in Graham \& Guzm\'an (2002) we demonstrate that if
the luminosity-profile shape ($M_B-n$) and the central surface
brightness-profile shape correlations ($\mu_0-n$) are universal, then
the change in slope observed in the $M_B-\langle\mu\rangle_e$ relation
(Figure 4a) for Es and dEs can be easily explained as a result of the
continuous change in the profile shapes (i.e., $n$) from dEs to
Es. This means that the observed change of slope in the
$M_B-\langle\mu\rangle_e$ relation is a mere artifact resulting from
the choice of parameters and does not have any physical implications
for a different formation mechanism at play in E and dE galaxies. The
key question is thus: are the $M_B-n$ ---or the $M_B-\mu_0$--- and
$\mu_0-n$ relations really universal (i.e., the same for E and dE
galaxies)?

	The $M_B-\mu_0$ and $\mu_0-n$ diagrams for our sample of 247
galaxies shown in Figures 4c and 4f reveal a well-defined, universal
correlation in both cases spanning over 10 magnitudes. However, bright
Es with $M_B<-20.5$ do deviate from the general trend defined by both
dE and E galaxies in the $M_B-\mu_0$ diagram\footnote{Note that the sample
of E galaxies with $M_B<-20.5$ from Faber et al.\ (1997) used in this
work did not have S\'ersic index  measurements and thus
could not be included in the $\mu_0-n$ diagram.}. All these bright E
galaxies are classified as ``core'' ellipticals by Faber et al.\ \
(1997). The simplest physical interpretation of this behaviour is that
both dE and E galaxies do follow the same universal correlations till
the onset of core formation in the brightest elliptical galaxies.  The
observed cores in these luminous elliptical galaxies are thought to
have arisen from the partial evacuation of the nuclear region by
coalescing blackholes (e.g., Ebisuzaki, Makino, \& Okumura 1991;
Makino \& Ebisuzaki 1996). Due to this, or whatever process(es) that
have reduced these galaxies central surface brightness profiles, the
high luminosity core ellipticals clearly depart from this relation.
Thus {\sl bright Es with $M_B<-20.5$ are the exception ---not the
rule--- to the universality in the structural correlations of
spheroidal systems} and current models of bright elliptical galaxies
built from the merging of fainter spheroidal components {\sl must not
assume that the scaling laws defined by the bright ellipticals apply
to the pre-merged components}. A full account of the results of this
study can be found in Graham \& Guzm\'an (2002), Guzm\'an et al.\  (in
prep.), and Matkovi\'c \& Guzm\'an (in prep.).

\acknowledgements

R.G. would like to thank the organizing committee for such an
excellent meeting to honor our friend Roberto, both from the point of
view of the outstanding scientific results presented at the meeting
and the superb atmosphere of camaraderie. As Roberto likes to say:
``there is nothing like doing good science among friends''.

\section*{Discussion}

\noindent
{\it Kauffmann:} Can you comment on the environmental dependence on
the structural parameters?\\

\noindent
{\it Guzm\'an:} The galaxy harassment model makes very specific
predictions about the environmental effect on the structure of dE
galaxies. For instance: (i) only the densest dEs will survive in the
inner part of the cluster; (ii) an indirect correlation between
density and metallicity (e.g., as a result of a mass-metallicity
relation) will create a gradient in metallicity of dEs with
clustercentric distance; (iii) since nucleated dEs will be more robust
than those without a nucleus, the fraction of nucleated dEs will
increase toward the cluster centre; (iv) nucleated dEs will show a
higher central velocity dispersion than non-nucleated dEs. \\

\noindent
{\it Moss:} Besides considering the proposed scenario of galaxy 
harassment for the origin of dwarf ellipticals have you considered mergers
of HII galaxies as their possible origin? \\

\noindent
{\it Guzm\'an:} This is a plausible alternative scenario. Indeed, an
evolutionary connection between HII galaxies and dE galaxies has been
proposed by several authors. However, to the best of my knowledge, such
an evolutionary scenario lacks a well-constructed theoretical model whose
predictions we can readily test with our dataset.\\

\noindent
{\it Zinnecker:} Do you think all the progenitors to nucleated dwarf 
ellipticals survived to the present day, and if not, where are these dense
nuclei now?\\

\noindent
{\it Guzm\'an:} If they do not, this would be indeed a most
interesting puzzle. The derived luminosities of the nuclei in our dE
sample are of the order of $10^6$-$10^7$ $L_\odot^B$. Perhaps, as you
argue in your talk, they may become the massive W-Cen-like globular
clusters in more massive galaxies. \\

\end{document}